\documentclass[runningheads]{llncs}

\global\long\def\ket#1{\left|#1\right\rangle }
\global\long\def\bra#1{\left\langle #1\right|}

\usepackage{ifr_math}
\usepackage{graphicx}
\usepackage{amsmath}
\usepackage{physics}
\usepackage{amssymb}
\usepackage{booktabs}
\usepackage{mathtools}
\usepackage{footnote}
\usepackage{acronym}
\usepackage{soul,color}

\acrodef{SERP}[SERP]{Search engine result page}

\begin{document}
\title{Reinforcement Learning-driven Information Seeking: A Quantum Probabilistic Approach}
\titlerunning{RL-driven Information Seeking: A Quantum Probabilistic Approach}
%
\author{Amit Kumar Jaiswal\orcidID{0000-0001-8848-7041} \and 
Haiming Liu\orcidID{0000-0002-0390-3657} \and
Ingo Frommholz\orcidID{0000-0002-5622-5132}}
\authorrunning{Jaiswal et al.}
%
\institute{University of Bedfordshire\\
Luton, United Kingdom\\
\email{\{amitkumar.jaiswal,haiming.liu,ingo.frommholz\}@beds.ac.uk}}
\maketitle              

{\let\thefootnote\relax\footnotetext{Copyright \textcopyright\ 2020 for this paper by its authors. Use permitted under Creative Commons License Attribution 4.0 International (CC BY 4.0). BIRDS~2020, 30 July 2020, Xi'an, China (online).}}

\begin{abstract}
Understanding an information forager's actions during interaction is very important for the study of interactive information retrieval. Although information spread in uncertain information space is substantially complex due to the high entanglement of users interacting with information objects~(text, image, etc.). However, an information forager, in general, accompanies a piece of information (information diet) while searching (or foraging) alternative contents, typically subject to decisive uncertainty. Such types of uncertainty are analogous to measurements in quantum mechanics which follow the uncertainty principle. In this paper, we discuss information seeking as a reinforcement learning task. We then present a reinforcement learning-based framework to model forager exploration that treats the information forager as an agent to guide their behaviour. Also, our framework incorporates the inherent uncertainty of the foragers' action using the mathematical formalism of quantum mechanics.




\end{abstract}
%
%
\section{Introduction}
Web searchers, in general, move from one webpage to another by following links or cues while keeping the consumed information~(intake) with itself without attaining a generalised appetite~(information diet) in possession of uncertain and dynamic information environments~\cite{pirolli1995information,chowdhury}. In general, the evolution of information patterns from user interaction keeps searchers in an information seeking process to not consume optimised information diet~(the information goal). So, there needs to be a mechanism that can guide the foragers during their search process in order to set a realistic information appetite.
User interaction is an important part of the search process which can enhance the search performance and the information foragers' search experiences and satisfaction~\cite{norbert,brennan,liu2010applying}. User action and their dynamics during search play an important role in changing behaviour and user belief states~\cite{Rwhite}. It has been recently demonstrated that action behaviour representations can be learned using reinforcement learning~(RL)~\cite{chandak} by extrapolating a policy in two components~-~action representation and its transformation. To effectuate the information foragers'~(or searchers'/users')~\cite{wittek1} cognitive ability during the search, we treat the searcher as an RL agent which follows Information Foraging Theory~(IFT)~\cite{pirolli99}, to understand how the users can learn in an ongoing process of finding information. Furthermore, the learning ability of the users can be signalled by the RL approach through giving a free-choice of search scenarios in an uncertain environment. For instance, the information seeker must optimise the trade-off between exploration by sustained steps in the search space on the one hand and exploitation using the resources encountered on the other hand. We believe that this trade-off characterises how a user deals with uncertainty and its two aspects~--~risk and ambiguity during the search process~\cite{wittek1}. Therefore the pattern of behaviour in IFT is mostly sequential. Risk and ambiguity minimisation cannot happen simultaneously, which leads to an underlying limit on how good such a trade-off can be. This lets the information foraging perspective of information seeking converge with the developing field of quantum theory~\cite{wittek1}. Moreover, web search engines enable their users to efficiently access a large amount of information on the Web, which in turn leads search users to learn new knowledge and skills during their search processes. When the users search to obtain knowledge, their information needs\footnote{we consider an IN is expressed by a query or series of queries} are mostly varied, open-ended, and seldom not clear at the start. Such types of search sessions generally span multiple queries and involve rich interactions, therefore our aim is to model such kind of information foraging process where the users' cognitive state changes during search. 

Due to its inherently complex and intense interactive nature, the effective and interactive  information foraging process is exigent for both the users and the search systems. Hence, our focus is to incorporate contextual semantic information in modelling the information forager with the usage of the mathematical framework of quantum theory, i.e.\ quantum probabilities based on geometry. Specifically, we propose a quantum-inspired reinforcement learning approach that (a) models the information foragers' behaviour, where action-selection~(or policy) is leveraged as an Actor-critic method~\cite{acritic} to enhance the agent's experience in a text query-matching task; (b) learns the policy where query representation is parameterised using quantum language models, with a focus on the interaction across multi-meaning words. 





\section{Related Work}
This section covers aspects of  reinforcement learning, Quantum theory in dynamic information retrieval~(IR), in particular, interactive information retrieval~\cite{piwowarski,ztang}, and Information Foraging theory.

\paragraph{Reinforcement Learning in Information Retrieval:} Humans' transfer of information to other animals is a common method of learning and interaction, which is generally called reinforcement learning. Reinforcement learning~\cite{sutton}~(RL) techniques are motivated by our sense of decision making in humans which appears to be biologically rooted. Within such biological roots~\cite{charnov}, when an information foragers' action ends up with a disadvantageous consequence (or negative payoff), such action will not be counted in the future; whereas, if his/her action leads to a successful consequence (or positive reward), it will happen again. User involvement in information searching is primarily a decision making (or action taking) process~\cite{duspink}, where users reflect identical RL features during this process. We will adopt RL models to manifest the mechanisms prevailing users' learning of information from searching. Previous work~\cite{whitebennett} found that a search system's information can be enriched to advance search intention and automate the difficult query reformulation by modelling the search context. Reinforcement learning is an important method that can let the system employ the search context and relevance feedback simultaneously. Also, this approach allows the system to deal with exploration (widening the search among different topics) and exploitation (moving deeper into generic subtopics) which has been supportive in information retrieval~\cite{bzhang,yseo}. Exploration and exploitation methods are usually employed in tasks associated with recommender systems or information retrieval, such as foraging strategies~\cite{eliassen}, recommendation~\cite{yyue} or image retrieval~\cite{balabanovic}. However, reinforcement learning is mainly used by search/retrieval systems~\cite{rdynamic}, which collect users' interests and habits over a continuous period, while in a specific search scenario the users in a given search session are more interested in the holistic improvement of the search results than relying on arbitrary future search sessions.

\paragraph{Quantum Theory and Information Retrieval:} Quantum Theory (QT) has been matured to reinforce the search potential by employing the mathematical formalism of quantum mechanics to information retrieval~\cite{van2004geometry}. The aim of introducing the QT formalism was to elucidate the implausible behaviour of micro-level search actions, which classical probability theory may not be able to model. Furthermore, it is an expressive formalism that can combine prominent probabilistic, geometric and logic-based IR approaches. The mathematical foundation of the Hilbert space formalism was introduced in~\cite{von2018mathematical} to apply this mathematical framework outside of Physics. We refer to events as a subset of a sample space of all potential events in classical probability theory, whereas in Quantum theory, the probabilistic space is geometrically defined, and the representation of it becomes an infinite set of angles and distance commonly named as an abstract vector space~--- or, more appropriately, a finite or infinite-dimensional Hilbert Space denoted by $\mathcal{H}$. Each and every event is depicted as a subspace of the Hilbert Space. To represent the $n$-dimensional vectors that compose a Hilbert Space, the Dirac notation is widely adopted, using \emph{ket} and \emph{bra} nomenclatures. More concretely, this means representing one given vector $\psi$ as $\ket{\psi}$ and its transposed view, $\transpose{\psi}$ as $\bra{\psi}$. Also, the vectors under consideration in a Hilbert space are usually unit vectors (their length is 1). A projection onto a subspace induced by a vector $\ket{\psi}$ is denoted by the operation resulting in a matrix $\ket{\psi}\bra{\psi}$. In this subspace, the vectors contained within are again normalised\footnote{there may be some vectors which are not necessarily normalised}, and the projection of events represented as vectors, again, is performed by the $\ket{\psi}\bra{\psi}$ operation. Unit vectors interpreted as state vectors induce a probability distribution over events (subspaces), and the product resulting from the mentioned operation is called \emph{density matrix}. We use so-called observables to perform a measurement of the outcomes (which are eigenvalues). 


The major similarity between quantum mechanics~(QM) and information retrieval~(IR) is understanding the interaction between a user~(the observer in QM) and the information object under observation~\cite{van2004geometry}. The core connection between QM and IR stems from the probabilistic features, where there is an observation of agreement for the preface of conditional probabilities allied with interference effects dominating to some contextual measure (cognitive, subjective character) when consolidating varied objects\footnote{https://www.newscientist.com/article/mg21128285-900-quantum-minds-why-we-think-like-quarks/}. In QT, we can represent user information needs with state vectors, and the query/observable, eigenvalues and the probability of obtaining single eigenvalues or objects as a measure of the  degree of relevance to a query~\cite{piwowarski}. Earlier QM was incorporated withing the RL algorithmic approach to generalise on filtering favourable user actions~\cite{qmlike}.

\paragraph{Information Foraging Theory:} Information Foraging theory~(IFT)~\cite{pirolli99} was developed to understand human cognition and their behaviours. IFT provides stipulated constructs adopted from optimal foraging theory which includes predators conforming to humans who seek for information~(or prey). It has three constructs, one of which delineates searches~(or \ac{SERP}s) in the user interface sections, referred to as \emph{information patches}; \emph{information scent} helps users make use of perceptual cues, such as web links spanning small snippets of graphics and text, consecutively to make their navigation decisions in selecting a specific link. The purpose of such cues is to characterise the contents that will be envisaged by trailing the links. Finally, \emph{information diet} allows users to narrow or expand diversities of information sources based on their profitabilities~(appetite).

Information Foraging is an active area of IR and information seeking due to its sound theoretical basis to explain the characteristics of user behaviour. IFT has been applied to model users' information needs and their actions using information scent~\cite{chi2001using}. However, it has been previously found that information scent can analyse and predict the usability of a website by determining the website's scent~\cite{chi2000scent}. Liu et al.~\cite{liu2010applying} demonstrated an IFT-inspired user classification model for a content-based image retrieval system to understand the users' search preference and behaviours by functioning the model on a wide range of interaction features collected from the screen capture of different users' search processes. Recent work~\cite{jaiswal,icbir} studied the effects of foraging in personalised recommendation systems by inspecting the visual attention mechanism to understand how users follow recommended items. Such user-item interactions can also be seen in query-level interactions i.e., in query reformulation scenarios where IFT- and RL-like models~\cite{iqac,rlqr} provide better explainability. 
\section{Information Seeking As Reinforcement Learning Task}
A searcher during the search process has to investigate several actions before selecting any of it, with unknown reward. They explore each result back and forth to estimate the optimal patch based on the reward. This scenario of information seeking can be interpreted as a reinforcement learning task where the search process, involving an agent to interact with the search environment, is cost-driven. Assessing positively rewarded actions~(from searcher's incurred costs) by the agent within an uncertain environment can potentially optimise the foragers' choice in finding the information. From an IFT perspective, positively rewarded actions can be drawn as exploitation whereas the available actions as exploration provided the information must be scattered between patchy environment. The fundamental aspect of reinforcement learning is to ``learn by doing with delayed reward'', which emerges as a major connection to information seeking (especially user interaction in IR and recommendation tasks) and it also interprets the foraging process of a searcher. The seeker's goal is to quickly locate a relevant patch~(document, image, etc). However, the information seeker has no prior knowledge of the rewards from assessed patches and they keep exploring each of it. The seeker interacts with the search system to explore which results in relevant information elicits the rewards distribution (information scent patterns) between information patches; often the access to patches with minimum reward can signify an optimal patch that the seeker has spent less time on for exploitation. An information seeker spending less time assessing each information patch leads to partially-relevant information about the seeking process that elicits the rewards distribution between the patches, and it gives rise to exploitation of a patch with less than the optimal rewards. Hence, the longer a seeker explores, he/she consumes near-accurate information about all of the patches but gives up the chance to exploit the most relevant patch for long. Understanding these operationalised scenario paves the way to model foraging behaviour in which user causes could be uncertainty, information overload, and confusion.    
\section{Quantum-inspired Reinforcement Learning Framework}
We outline the proposed reinforcement learning approach to model the forager's action during an information seeking scenario where the task is to match a query for a given document in which the forager actions are queries.
An agent interacts with its search environment characterised by a patchy distribution of information to find an optimal foraging strategy to maximise its reward. The forager's environment provides a fixed setting of optional information sources. Moreover, the forager has the choice to add a distinctive type of information patch into their diet. However, the distribution of distinctive information patches may consist of information which the forager could likely not consume due to counterfactual situations in making decision amongst which patch (let us say document D1 and D2) contains certain information. In our framework, we consider the environment to be uncertain with dynamic parameters throughout a forager's search trail. The forager finds it difficult to differentiate patches and exploits experience to learn the environment. The increasing learning makes it complex at the dynamic and cognitive level where the forager's pursuit is to locate most relevant documents. 

We use the \emph{Actor-critic} policy gradient method~\cite{acritic} which inherently models such dynamics due to the forager's sequential behaviours that generate a continuous state representation. A forager's action~(or state) can be described with the quantum superposition state and the corresponding updated state vectors, based on the respective interaction, can be achieved by random observation of the simulated quantum state based on the collapse principle of quantum measurement~\cite{van2004geometry}. The probability of such an \emph{action state vector} can be obtained by the probability amplitude which will be updated in parallel based on reward. This gives rise to new internal aspects in traditional RL algorithms which are policy, representation, action~(in parallel) and operation update.

The quantum measurement decision process of a forager in selecting a document (the action) while seeking is ambiguous and uncertain~\cite{wittek1}. In such situation, an observable describes possible actions (documents or information patches to select) and can be represented as~($\hat{O}$) with a base set containing $\ket{0}$ and $\ket{1}$ which corresponds to the two state vectors of $\hat{O}$. The measurement of a quantum system on the observable~($\hat{O}$) in a corresponding superposed quantum state~($\ket{\psi}$) refers to a measurement in superposition state. When making a measurement in state $\ket{\psi}$, the quantum state would collapse into one of its basis states $\ket{0} or \ket{1}$. However, one cannot obtain a prior with certainty whether it will collapse to either of these states. The only information this quantum system can provide is $\ket{0}$ will be measured with probability $\abs{\alpha}^2$ or $\abs{\beta}^2$ as the probability to measure $\ket{1}$, where $\alpha$ and $\beta$ represent the respective probability amplitudes. 

\begin{figure}[tb]
\includegraphics[width=\textwidth]{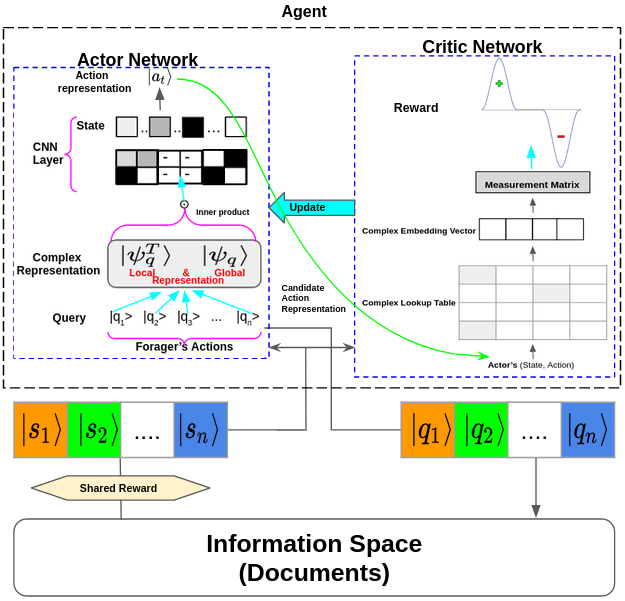}
\caption{Our proposed framework: Quantum-inspired Reinforcement learning-driven model for information seeker~(in a semantic query matching task)} \label{fig:model}
\end{figure}
We present a \emph{quantum-inspired reinforcement learning~(qRL)} framework for information seeking under dynamic search scenarios. The schematic architecture of qRL is shown in Fig.~\ref{fig:model}. qRL has two main components, an Actor-critic~\cite{acritic} based network to represent the RL agent which jointly encodes state and action spaces, and the information space known as environment containing documents. The Actor-critic components of an RL agent have their constructs subscribed via the Hilbert space formalism of Quantum theory~\cite{van2004geometry}. 

Our framework is applicable to matching tasks, in particular, \emph{semantic query matching} where candidate queries (extracted/predicted queries from the document) with the original document will be matched in a semantic Hilbert space (SHS)~\cite{SHS}. An SHS is a vector space of words, where words in combination involve a linear/non-linear formation of amplitudes and phases, delineating various level of semantics of combined words. In the SHS, a word $w_i$ is represented by a base vector $\ket{w_i}$. Semantics of combined words are represented by superpositions of word vectors, encoded in the probability amplitudes of the corresponding base vectors.
\subsection{Preliminaries}

The standard reinforcement learning is based on a finite state, discrete time Markov decision process~(MDP) composed of five components: $s_{t}, a_{t}, p_{ij}(a), r_{i,a}$ and $C$, where $s_{t}$, the state at time $t$, delineates $a_{t}$~the action at a specific time for a given state; $p_{ij}(a_t)$ is the probability of state transition~(from state $s_{t}$ to $s_{t+1}$ via action $a_t$ for all $t\in (i,j)$), \emph{r} is a reward function where $r: \Gamma\longrightarrow \mathbb{R}$ with $\Gamma = \{(i,a)~|~i\in s_t,~a\in a_{t}\}$, and $C$ is an objective function.

In the following discussion we utilise tensor spaces. The notation in Table~\ref{tab:not} follows those in~\cite{ppooling,tensor}. The fabric of our framework, i.e.\ the underlying Hilbert space $\mathcal{H}$, is similar to the Tensor Space Language Model described in~\cite{tensor}. Here, the base vectors $\{\ket{\phi_i}\}_{i=1}^n$ of our $n$-dimensional space\footnote{The Hilbert space can be over the real or complex field, i.e.\ $\R^n$ or $\C^n$; we are assuming $\R$ for the further discussion} are term vectors, either one-hot vectors or word embeddings. Any word vector $\ket{w}$ can be written as linear combination of the base vectors, i.e.\ $\ket{w} = \sum_{i=1}^n \alpha_i\ket{\phi_i}$ with $\alpha_i \in \R$ (or $\C$ in the complex case) as coefficient. 
\small
\begin{table}
  \centering
  \caption{Notations used in Reinforcement Learning Constructs, following~\cite{ppooling,tensor}. $b_i$ depicts the dimension of orthonormal basis of Hilbert space, $\otimes$ depict the tensor product, R depicts the rank of $\mathcal{G}$ and $\mathcal{L}$ has n-order tensor of rank 1.}
  \label{tab:not}
  \begin{tabular}{lcp{5cm}}
    \toprule
    \textbf{Notation} & \textbf{Interpretation} & \textbf{Description}\\
    \midrule\midrule
    $\alpha_{i,b_i}$ & $b_i\in\{1,..., k\}$ & Probability amplitude\\ \hline
    $\ket{\phi_{b_i}}$ & Semantic meaning & Basis vector~($n$ (word vectors) or $k^n$ dimension for tensor product of basis vectors)\\ \hline
    $\ket{w_i}$ & $\sum_{b_i=1}^{k}\alpha_{i,b_i}\ket{\phi_{b_i}}$ & Word state vector \\ \hline
    $\ket{q_i}$ & $\ket{w_1}\otimes\ket{w_2}...\otimes\ket{w_n}$ & Query state vector \\ \hline
    $\ket{\psi_{q}^{T}}$ & $\sum_{b_1,\ldots,b_n=1}^{k}(\underbrace{\prod_{i=1}^{n}\alpha_{i,b_i}}_{\mathcal{L}_{b_i,\ldots,b_n}}\ket{\phi_{b_i}}\otimes...\otimes\ket{\phi_{b_n}})$ & Local representation~($\mathcal{L}$ is a $k^n$ dimensional tensor)\\ \hline
    $\ket{\psi_q}$ & $\sum_{b_1,\ldots,b_n=1}^k \mathcal{G}_{b_1b_2...b_n}\ket{\phi_{b_i}}\otimes...\otimes\ket{\phi_{b_n}}$ & Global representation of combined meanings/patches \\ \hline
    $\mathcal{G}$ & $\sum_{r=1}^{R} w_r\cdot e_{r,1}\otimes e_{r,2}\otimes...\otimes e_{r,n}$ & Probability amplitude (semantic space of meaning)\\ \hline
    $\bra{\psi_{q}^{T}}\ket{\psi_q}$ & $\sum_{b_1,\ldots,b_n=1}^k\underbrace{\mathcal{G}_{b_1\ldots b_n}\times\prod_{i=1}^{n}\alpha_{i,b_i}}_{Probability~amplitudes}$ & Projection of the global representation to the local representation of a query \\ \hline
    State & $\prod_{i=1}^{n}\sum_{b_i=1}^{k} e_{r,i,b_i}\cdot\alpha_{i,b_i}$ & Actor network state module (product pooling layer~\cite{ppooling})\\ \hline
    $\ket{a_t}$ & $(\ket{a_1},\ket{a_2},...,\ket{a_R})^T$ & Output of the Actor network \\
  \bottomrule
\end{tabular}

\end{table}

The overview of the RL process which possesses the Markov decision process is as follows:
\begin{description}
    \item[Agent:] In general, an agent acts as a controller within an information environment and is the one executing actions. In our framework, the RL agent is a forager (information seeker) available to the search environment (documents) delivering queries as actions, where the action is chosen based on the Actor-critic network~(or the Policy network).
    
    \item[Action:] An action $a_t$ in the Web search scenario conforms to a query that searchers utilise to express their information need with the aim to either retrieve a document as an outcome of the query or continue the search process (exploratory search)~(a formal representation of the user action is shown in Table~\ref{fig:model}). In our framework, the forager (or searcher) action is to match a candidate query~$\ket{q}$ (generated after inputting a set of queries) from document~$D$ to delineate $\ket{q_{rD}}$, where $\ket{q_rD}$ refers to a query state vector that represents the most optimal query for the selected document $D$ given a positive/optimal reward~($r$). A candidate query is an outcome generated from the  Actor network given the forager set of input queries.
    
    \item[State:] A state $s_t$ delineates the positive historical interaction of the forager with the search environment. In our framework, the Actor network has its \emph{state} encoded by the product of the probability amplitudes of global-local projection~$\bra{\psi_{q}^{T}}\ket{\psi_q}$ (of word meanings) for all words of a query\footnote{Please note the superscript $T$ does not mean the transpose in this context but denotes tensor product subscribed local/global representation (see Table~\ref{tab:not})}. We refer to the state representation defined by the product pooling method. 
    \item[State Transition:] The state representation describes the positive historical interaction of a forager. The transition among the states can be computed from the user's feedback. Our framework uses a convolutional neural network which has its convolution based on a state vector that encodes the historical interaction of the forager in finding the match of a query.
    \item[Policy:] Policy is a strategic mechanism which represents the probability of a forager's action under a certain state. Our framework's policy network is stochastic, and we employ the Actor-Critic RL method~\cite{acritic} (Fig.~\ref{fig:model}) which assists the forager actions in the Actor network with a optimal policy value generated from the Critic. Thus, the Actor network estimates the probability of a forager action, and the Critic network gets the optimal value and updates it. The policy network is modelled as a probability distribution over actions and hence it is stochastic. 
    \item[Reward:] Reward~($r(s,a)$) in reinforcement learning is the success value of an agent's action $a$. This success value in information retrieval is interpreted in terms of the relevance judgement score~\cite{ztang}. Our framework process to receive reward values for the Critic network which inputs a pair of~(state, action) that provides to the Actor network as an optimal reward for a given action which judges and scores the actions of the agent (or forager).
\end{description}

\subsection{Our Proposed Framework}
This fundamental RL definition is of utmost importance for proposing quantum-inspired reinforcement learning constructs. Following the quantum probability concepts, below are the constructs as follows (please also refer to Figure~\ref{fig:model}):\par
\textbf{Actor Network:} An Actor-critic~\cite{acritic} method refers to as policy gradient mechanism, where the Actor network for a given forager~(or information seeker) in a particular state~$\ket{s_t}$ outputs an action~$\ket{a_t}$. This network inputs user queries (forager's actions), where these queries~$\ket{q_1}, \ket{q_2},...,\ket{q_n}$ or a set of textual descriptions~(which collectively form a document) form the local and global representations so as to model the inter-related interaction between words. Inspired by the notion of quantum theory, we employ the interpretation of wave function~(due to the importance of word positions~\cite{wordwave})~$\ket{\psi}$ as a state vector that can be explicated in RL constructs. 

The Actor network inputs query state vectors $\ket{q_1}\ldots, \ket{q_n}$, where each word in a query is treated as a state vector~$\ket{w_i}$ and every word has a unique basis vector~$\ket{\phi_{b_i}}$ that provides a generic semantic meaning with an associated  probability amplitude. The speciality of a basis vector is that it can lead to a different meaning if interpreted severally across it. Then, we apply our framework in a semantic query matching task by a real-valued representation of queries by means of local and global distribution so to allow such intermittent basis vectors that perceive the interaction between the meaning of different words. Hence, the wave function description of a query~$\ket{q_i}$ can be depicted using the tensor product of words as $\ket{\psi_{q}^{T}} = \ket{w_1}\otimes\ket{w_2}...\otimes\ket{w_n}$. A word dependency can be seen by expansion of tensors as $\ket{\psi_{q}^{T}}$ = $\sum_{b_1,\ldots,b_n=1}^k \mathcal{L}_{b_1\ldots b_n}\ket{\phi_{b_i}}\otimes...\otimes\ket{\phi_{b_n}}$, where $\mathcal{L}$~(the value is shown in Table~\ref{tab:not}) depicts the allied probability amplitude of $k^n$ dimensional tensor in which it has the respective basis vectors~$\ket{\phi_{b_i}},...,\ket{\phi_{b_n}}$ representing the meaning of the corresponding query. This tensor-based query representation is a local representation as a tensor with rank 1 actually delineates the local distribution of a query~\cite{tensor}. For words that are unseen in a query or compound meanings we need a global representation of them provided by  a collective set of basis states (or vectors). A state vector (i.e., wave function of a query) to describe such a global representation is $\ket{\psi_q} = \sum_{b_1,\ldots,b_n=1}^k\mathcal{G}_{b_1\ldots b_n}\ket{\phi_{b_i}}\otimes...\otimes\ket{\phi_{b_n}}$. This wave function delineates a semantic embedding space of $n$ uncertain word meanings of a given query. The local and global representation differs in terms of their corresponding probability amplitudes i.e., $\mathcal{L}$ and $\mathcal{G}$, in which the probability amplitudes of the \emph{global distribution} will be trained on a large collection of previous queries whereas the probability amplitudes of \emph{local distribution} relates only to the input query. To compute the probability amplitudes among words from the input query (local representation) and unseen words generated from the global representation, we perform the inner product~$\bra{\psi_{q}^{T}}\ket{\psi_q}$ of both representations that disentangle the interaction among it. The value of the projection is shown in Table~\ref{tab:not}. We use a convolutional neural network~(CNN) to learn the obtained higher-dimensional tensor~$\mathcal{G}$~(value shown in Table~\ref{tab:not}), where tensor rank decomposition can be used to decompose it (among other methods such as generalised singular value decomposition) and the decomposed unit vector~$e_{r,n}$ with each rank 1 tensor of weight coefficient~$w_r$. The unit vector is $k$-dimensional and the set of vectors~$e_{r,n}$ acts as a subspace of tensor~$\mathcal{G}$. The CNN inputs a query state vector with a convolution filter composed of the projection (inner product) among the $\ket{q}$ and the decomposed vector, which makes the CNN trainable. Then, the state representation (actor's state in Table~\ref{tab:not}) performs the product of all mapped unit vectors (from~$\mathcal{G}$) for all the sub-words of a query. After all these operations, the Actor network yields an action state vector~$\ket{a_t}$ (action $a_t$ at time $t$) to depict a set of matched words.\\
\\
\textbf{Critic Network:} The Critic network of the qRL framework is based on a quantum-like language model parameterised CNN which inputs the generated state and the candidate action~$\ket{a_t}$ from the Actor network. The output of the Critic network is a scalar value or value of the Q-function~\cite{sutton}. The reward values~$R_e\in[-1,0,1]$ reflect the ability of the candidate action generated by the Actor network. The significance of the reward value represents the probability of designating the correct label to action i.e, the multi-class classification of queries to match among documents will be used to update the reward. Rewards (or classification labels) are categorised as -1 for a mismatched query which has negative word polarity (leading to a compound meaning). For instance, "dogs chase cats" and "dogs do not chase cats" contribute to a compound meaning itself but in an opposite sense. We tend to consider that a word renders the entire polarity of a query, provided to which new word it associates with. A realistic example of this hypothesis can relate to one of our framework's main constructs i.e., $\ket{q}$ which is a state vector equal to the tensor product of possible words, where the word coefficients (i.e., probability amplitudes) of basis vectors can be altered to derive a new query giving rise to a compound meaning. The negative word polarity example is an actual scenario of it. Positive and zero rewards are classed as matching and partially matching for queries.\par
In the Critic part, the concatenation of the actor's state and candidate action is performed using one-hot encoding in which the query is passed via a complex-valued lookup table, where each word in their own superposition state is encoded into a complex embedding vector~\cite{qce}. Then, a measurement is performed using the square projection to compute the query density matrix from the complex embedding vectors. The probability of a measurement can be estimated using the Born's postulate for a given query state~$\rho$ (a density matrix) which is $\mathbf{p} = \Tr(P\rho)$, where $\mathbf{p}$, $P$ and $\Tr$ represents the class of the query, projection matrix, and trace of a matrix, respectively. The density state~$\rho=\sum_{i=1}^{n}\beta_i\ket{w_i}\bra{w_i}$ of a query is perceived as the word states in combination, provided that the density matrix~($\ket{w_i}\bra{w_i}$) reflects a word~($w_i$) in superposition state (in this case $\sum_{i=1}^{n}\beta_i=1$). The generated query density matrix has its diagonal and non-diagonal entries as real-valued and complex nonzero values, and both type of entries inherently inform about the distribution of semantic and contingent meanings. We adopt the interpretation of complex phase introduced in~\cite{qce} to compute the sentence density matrix which has word senses as positive, neutral, and negative. The reward is estimated using such interpretation from the measurement matrix. A pictorial representation of the Critic network is shown in Fig.~\ref{fig:model}.

In brief, the Actor-Critic policy network helps suffice the number of components with respect to traditional reinforcement learning. Also, the Agent part of the framework acts as a controller for the user in the same way Information Foraging mechanisms possess to a searcher. IFT helps a searcher through suggesting an optimal foraging path via information scent, and here in the framework the Critic network informs/updates the Actor with a value (reward) for a certain action that is positively rewarded. Hence, our framework meets foraging in certain regards~(such as information seeking behaviour assessed as foraging and inherently as RL task).\\

\textbf{Rewards:} The forager aim is to identify the relevant match~(or a perfect match) of a query (or patch) for the clicked/selected document that can be perceived as its reward. However, our framework's reward function is designed in a way to guide the forager on how to perceive the document information and draw the most relevant match~(patch). Also, the reward value is discrete as it revolves around -1, 0, and +1. The definition of reward in reinforcement learning~\cite{sutton}\footnote{Rewards can be normalised to generate outcomes in reinforcement learning~\cite{sutton}} resembles a certain analogy of information scent, which is a measure of utility and results in two types of information scent score~-- a scalar value and the probability distribution of scent patterns~\cite{iqac}. In RL, the perspective of value distribution of received reward by an agent can depict the analogous nature of information scent patterns. Hence, an explainable approach of reinforcement learning-based rewards using the IFT-based model of information scent can give further intuition to negative rewards. Information scent can be interpreted as the perceived relevance of rewarded actions defined as positive and negative scent values. The physical meaning of positive and negative information scent scores are that the forager accumulates rich information along the path he/she foraged to locate the relevant information, and the unhealthy consumption of information reckon searcher negative towards the search environment which leads them to give up the information world~(or RL environment) or task itself.
\\

\textbf{Update Probability Amplitude/Policy:} To update the probability amplitude in the Actor network, the important part is to measure the actions for some certain states which on collapse will give rise to the occurrence probability of the norm of state vector for the particular candidate action, which later will execute the Actor network. The more we record the experience and learning of each action~(even erroneous action), the probability amplitude becomes more informative. We know that the action $\ket{a_{t}}$ is the tensor product of all possible words and to calculate one user action (i.e.\ $\ket{a}$) from it can be possible while interacting with changes in probability amplitudes for the combined meaning.

%
\section{Conclusion and Future Work}
In this paper, we propose a mathematical framework of reinforcement learning inspired by the Hilbert space formalism in Quantum theory. The framework models the learning process of forager actions in a semantic query matching task given the search environment is patchy. The core of our framework is to characterise a forager with very little or unclear information about their search pattern, unclear or evolving information need and features. Also, no information about how a forager makes their trail (initially the information scent is unknown and emanates as it follows via distinct cues) choice during finding information and the amount of information they consume in real-time interaction with the search system. Apart from this, the major trade-off situation of exploration and exploitation in the foraging process makes the process of understanding about the forager's search actions complex. To tackle such a complex process of dynamic action for a state and vice-versa, we adapt the Actor-critic reinforcement learning method as a policy network, in which the actor network is continuously informed about the generated action from the critic network. The framework subscribes to the quantum probability constructs to model by the representation of forager search actions and states. Quantum theory has been earlier applied in the area of information seeking~\cite{wittek1}, but representing and measuring actions of each state is a challenging scenario due to the continuous update of state-action in parallel, so using the Actor-critic reinforcement learning method paves the way to influence learning and representation mechanisms; many complex IR problems could be interpreted appropriately in a new way within such an inclusive framework. In the future, we intend to evaluate this framework for certain IR tasks. 
\vspace*{-4mm}
\section*{Acknowledgements}
\vspace*{-2mm}
This work is part of the Quantum Access and Retrieval Theory (QUARTZ) project, which has received funding from the European Union's Horizon 2020 research and innovation programme under the Marie Sklodowska-Curie grant agreement No 721321.
\vspace*{-4mm}
\bibliographystyle{splncs04}
%

\end{document}